\documentclass[11pt,onecolumn]{aastex}   
\usepackage{natbib,emulateapj5,psfig}

\newcommand{\be}{\begin{equation}}
\newcommand{\ee}{\end{equation}}
\newcommand{\bea}{\begin{eqnarray}}
\newcommand{\eea}{\end{eqnarray}}
\newcommand{\eqlab}[1]{\label{eq:#1}}
\newcommand{\eqref}[1]{\ref{eq:#1}}
\newcommand{\pder}[2]{{\partial #1 \over \partial #2}}

\newcommand{\lya}{{Ly$\alpha$}}
\newcommand{\rii}{R_{\rm II}}
\newcommand{\DeltaD}{\Delta\nu_{\rm D}}
\newcommand{\Rstar}{R^{*}}
\newcommand{\Rcorr}{R_{\rm c}}
\newcommand{\Diff}{D }
\newcommand{\TR}{T_{\rm R}}
\newcommand{\bfr}{{\bf r}}
\newcommand{\bfn}{{\bf n}}

\shorttitle{Resonance Line Scattering}
\shortauthors{Rybicki}

\begin{document}

\title{Improved Fokker-Planck Equation for Resonance Line Scattering}
\author{George B. Rybicki}
\affil{Harvard-Smithsonian Center for Astrophysics}
\affil{ 60 Garden Street, Cambridge, MA 02138 USA, grybicki@cfa.harvard.edu}

\begin{abstract}

A new Fokker-Planck equation is developed for treating resonance line
scattering, especially relevant to the treatment of \lya\ in the early
universe.  It is a ``corrected'' form of the equation of Rybicki
\& Dell'Antonio that now obeys detailed balance, so that the approach 
to thermal equilibrium is properly described.  The new equation takes
into account the energy changes due to scattering off moving
particles, the recoil term of Basko, and stimulated scattering.  One
result is a surprising unification of the equation for resonance line
scattering and the Kompaneets equation.  An improved energy exchange
formula due to resonance line scattering is derived.  This formula is
compared to previous formulas of Madau, Meikson, \& Rees (1997) and
Chen \& Miralda-Escud\'e (2004).

\end{abstract}


\section{Introduction}

The scattering of radiation in resonance lines plays an important role
in many parts of astrophysics, including stellar atmospheres, diffuse
nebula, and active galactic nuclei.  In the early universe, resonance
scattering in the hydrogen \lya\ line has been shown to be an crucial
process for understanding the state of the primordial gas during the
recombination era \citep{Peebles68,Zeldovich69}.  It can also can
play a critical role in determining the populations of the fine
structure levels in the 1$S$ level of hydrogen
\citep{Wouthuysen52, Field58, Field59}, and thus affect the formation
of the 21cm line, observations of which promise to be an important
source of information about the early universe \citep{Hogan79, MMR97,
Tozzi00}.  It may also play a role in direct heating of the gas
\citep{MMR97, Chen04}.

The theoretical description of resonance line scattering is based on a
{\em redistribution function} $R$, which gives the probability
that an initial photon state will be scattered into some final photon
state.  The level of description of these photon states depends on the
problem treated.  For this paper we confine our attention to problems
where the radiation field is sufficiently isotropic that only the
angle-averaged form of the redistribution function needs to be used,
and also that the radiation can be considered unpolarized.  The
redistribution function then depends only on the intitial $\nu'$ and
final $\nu$ frequencies of the scattered photon, that is,
$R(\nu,\nu')$.  

For \lya\ and other resonance line scattering in low density media,
coherent scattering in the atom's rest frame from a natural (Lorentz)
profile is appropriate.  When one accounts for the Doppler effect due
to atoms with a Maxwellian velocity distribution, a scattered photon
will have its frequency decreased or increased depending on the
components of the atomic velocity along the initial and final photon
directions.  The appropriate redistribution function under these
conditions was first derived by \citet{Henyey41} and, following
\citet{Hummer62}, is usually denoted $\rii(\nu,\nu')$.  For many cases in
astrophysics, $\rii$ captures the dominant physics for resonance line
transfer.

The solution of line transfer problems using a full redistribution
function $\rii$ is not trivial, since all frequencies are coupled
together.  However, there is an important class of problems where an
approximate formulation can effectively used, namely, when the
radiation field is sufficiently smooth on the scale of the Doppler
width of the line.  Then it is possible to derive a {\em
Fokker-Planck} (F-P) type of transfer equation, where the
redistribution is taken into account by a second-order differential
operator over frequency space.  An equation of this type was first
given by \citet{Unno52}, and improvements were made subsequently by
\citet{Harrington73}, \citet{Basko81}, and by
\citet[][hereafter RD]{RD94}. 

Another effect on redistribution is the loss of photon energy during
scattering due to the recoil of the atom.  This is completely
analogous to the ordinary Compton effect, but is smaller by the
ratio of electron to atomic mass.  The recoil effect was first
discussed by \citet{Field59} for the case of resonance scattering with
zero natural line width \citep[$R_{\rm I}$ in the notation of ][]{Hummer62}.
\citet{Adams71} qualitatively considered recoil for $\rii$ in the context 
of \lya\ scattering in the neutral hydrogen of the galactic disk.
\citet{Basko81} derived the appropriate generalization of the $\rii$ 
including recoil redistribution function and showed that it led to a
simple additional term in the F-P formulation.  \citet{RD94}
showed how the Basko term could be included in their F-P equation
and how it affected its solutions.  

In all the cases treated by Adams, Basko, and RD the recoil term did
not seem to be important, at least for its effect on the radiation
field itself.  However, \citet{MMR97} argued that the atomic recoil in
\lya\ scattering could be an important heating mechanism for the intergalactic
medium before reionization, basing their analysis on a simple formula
that included recoil but no other effect.  This was also investigated by
\citet{Chen04}, who found a much lower heating rate using the RD
F-P equations with Basko term. 

A strong motivation for the present paper is to understand better the
difference between the results of \citet{MMR97} and \citet{Chen04}
regarding energy exchange.  It seems clear that
\citet{Chen04} must be at least partially right: if the radiation field
were thermal at the same temperature as the gas, then no net energy
exchange should occur, so that the simple formula of \citet{MMR97}
cannot be of general validity.  On the other hand, the formulation of
\citet{Chen04} did not take detailed balance fully into account, so
the matter deserves further investigation.

Another motivation for this paper is the issue of the 21cm spin
temperature and how it depends on \lya\ transfer.  According to the
Wouthuysen-Field effect \citep{Wouthuysen52,Field58,Field59} the radiation field
near \lya\ comes into thermal equilibrium with the kinetic temperature
of the gas, and this \lya\ field then imposes that same spin
temperature on the 21cm line.  Any transfer equation, either using a
full redistribution function or through a F-P approximation,
should correctly incorporate this thermal requirement, since small errors
in calculation could have large effects on predictions of 
21cm radiation.

It should be obvious that an accurate determination of energy exchange
rates and 21cm spin temperatures is critically dependent on being able
to formulate the interaction of of radiation and matter in a
thermodynamically correct way.  \citet{DW85} showed that this will be
true if the redistribution functions satisfy certain {\em detailed
balance} relations.  However, no known form of $\rii$ satisfies
these relations exactly, so \citet{DW85}, in their investigations of
the recombination era, found a method of ``correcting'' the usual
$\rii$, such that its values did not change very much, but enough to
satify detailed balance.  
\citet{DW85} applied their correction method to the redistribution functions
directly and solved the equations numerically; they did not consider
how this might change any associated F-P method.

A review of the basic equations is given in \S\ref{basic}.
In \S\ref{detailed} we introduce a simple correction scheme, different
from that of \citet{DW85}, which can be used to produce a
redistribution function satisfying detailed balance, and we use this
correction scheme to derive a new, corrected F-P equation.  This F-P
equation incorporates the RD diffusion term, the Basko recoil term,
plus a new term that describes stimulated scattering and another that
accounts for the correct phase space factors; acting together, these
terms ensure detailed balance and the approach to the proper thermal
radiation field.  In \S\ref{Kompaneets} we demonstrate the close
analogy between the new corrected F-P equation and the Kompaneets
equation.  In \S\ref{exchange} we use the new F-P equation to derive an
energy exchange rate formula.  This will be compared to the the formulas of
\citet{MMR97} and of \citet{Chen04}.

\section{Basic Equations}\label{basic}

We shall consider the time dependence of the radiation field in a
homogeneous and isotropic medium.  This idealized problem allows one
to develop equations applicable to more complex situations under the
isotropic scattering approximation, much as is done for the analogous
Kompaneets equation.  When dealing with scattering problems, it is
often easier to use intensities based on photon numbers rather than
energies, and this is done here.  Because of the isotropic assumption,
the radiation is determined by the mean intensity $J(\nu)$. 

For simplicity of presentation, we shall write all of our transfer
equations for the special case of the time dependence of a
homogeneous, static medium, very much as in the spirit of the usual
Kompaneets equation \citep{Kompaneets57,RL79}.  One should interpret
the right hand side of such equations as ``emission minus absorption''
terms, which can be incorporated into other transfer equations as
needed.  In particular, we do not include a term accounting for the
expansion of the medium, which, if needed, should be included on the
left hand side as part of the transfer operator.  For cases that
involve spatial transfer, one can also use the results here in an
``isotropic scattering approximation'' by appropriate modification of
the absorption terms, that is, by adding back the isotropic absorption
term and subtracting the proper anisotropic absorption term.

For many applications the process of stimulated scattering is not
important, and most of the discussions of resonance line scattering
have explicitly or implicitly neglected it.  In that case the time
dependence of the radiation is then governed by the transfer equation,
\be
   {1 \over c\chi}\pder{J}{t} = \int \left[ 
  R(\nu,\nu')J(\nu')-R(\nu',\nu)J(\nu)
     \right] d\nu' = -\varphi(\nu)J(\nu)+\int 
  R(\nu,\nu')J(\nu')\, d\nu'.    \eqlab{1}
\ee
Here $R(\nu,\nu')$ is the $\rii$ redistribution
function and  $\varphi(\nu)$ is the Voigt line profile function, related
to the redistribution function by,
\be
     \varphi(\nu) = \int R(\nu',\nu)\, d\nu', \eqlab{2}
\ee
and which is normalized to unity,
\be
      \int_0^{\infty} \varphi(\nu)\, d\nu =1.    \eqlab{2.1}
\ee
The quantity $\chi$ is given by
\be
      \chi = {h\nu_0 \over 4\pi} N_1 B_{12},         \eqlab{3}
\ee
where $\nu_0$ is the line center frequency and $B_{12}$ is the upward
Einstein $B$-coefficient for the resonance transition.

When stimulated scattering is important, it can be taken into account
through additional factors in the transfer equation $(1+n)$, where
$n(\nu)=c^2J(\nu)/2\nu^2$ is the photon occupation number,
\be
   {1 \over c\chi}\pder{J}{t} = \int \left\lbrace 
  \left[ 1+ n(\nu) \right]R(\nu,\nu')J(\nu')
 -\left[ 1+ n(\nu') \right]R(\nu',\nu)J(\nu)
     \right\rbrace d\nu'.    \eqlab{4}
\ee
It is important to note that the same redistribution functions appear in
both equations (\eqref{1}) and (\eqref{4}), that is, they are independent
of whether stimulated scattering takes place or not.  Therefore, it
is possible to determine the properties of the redistribution functions
for cases without stimulated processes, and then use them in equation
(\eqref{4}) when necessary.

Both equations (\eqref{1}) and (\eqref{4}) conserve photons, as can be
proved by integration over $\nu$ and interchanging $\nu$ and $\nu'$ in
one of the two terms on the right.  As such, photon conservation is
built into the transfer equations themselves, and does not depend on a
particular symmetry or other property of the redistribution functions.
This is as it should be, since scattering does not change photon number.

The difficulty of solving transfer
problems using the full redistribution functions has led to some
simplified formulations that are adequate under special circumstances.
One important fact is that the redistribution function is
significantly nonzero only for frequencies $\nu$ and $\nu'$
that do not differ by many Doppler widths.  In many cases of interest 
the radiation field in the neighborhood of
line center is sufficiently smooth such for a given value of $\nu$,
the relevant values of $J(\nu')$ can be represented
by a Taylor series up to second order in the difference $(\nu'-\nu)$.
This makes it possible to replace the frequency integrations on
the right hand side of the transfer equation with an equivalent
second-order differential operator.  The resultant transfer equation
is known as a {\em Fokker-Planck} (F-P) equation.  

There have been a number of Fokker-Planck formulations for the $\rii$
redistribution problem given in the literature.  They fall into two
categories: first there are ones that are based on the standard $\rii$
redistribution function \citep{Henyey41,Hummer62}, which consider the
spreading of radiation on scattering but neglect recoil.  These
formulations give an expression for the integral of the redistribution
function and the intensity by means of a differential operator,
\be
 \int R(\nu,\nu') J(\nu')\,d\nu'  = \varphi(\nu) J(\nu)
 +{\DeltaD^2\over 2} \pder{}{\nu}
    \left[ \Diff(\nu) \pder{J(\nu)}{\nu} \right], \eqlab{5}
\ee 
where $R=\rii$.  Here,
\be
     \DeltaD= \left({2kT\nu_0^2 \over m_{\rm a} c^2}\right)^{1/2}, \eqlab{18.2}
\ee
is the usual Doppler width and $D(\nu)$ is the ``diffusion''
coefficient, essentially a second moment of the redistribution
function.  This implies the F-P transfer equation (without
stimulated processes),
\be
 {1 \over c\chi}\pder{J}{t} ={\DeltaD^2\over 2} \pder{}{\nu}
    \left[ \Diff(\nu) \pder{J(\nu)}{\nu} \right]. \eqlab{18}
\ee 

The different F-P formulations differ in what they take for
$D(\nu)$.  The first such formulation by \citet{Unno55}
\citep[see also][]{Harrington73} took
\be
      D(\nu) = {\Gamma \over 4\pi^2 |\nu-\nu_0|^2},  \eqlab{18.3.1}
\ee
which is an asymptotic result for the second moment, strictly 
applicable only in the wings of the line.  Basko used a non-singular
form of (\eqref{18.3.1}), 
\be
  D(\nu) = {\Gamma/4\pi^2 \over (\nu-\nu_0)^2 + (\Gamma/4\pi)^2},  
     \eqlab{18.3.2}
\ee
equal to the natural line profile, but did not justify this choice.  
\citet{RD94} rigorously caculated the second moment of $\rii$ using 
the full $\rii$ redistribution function.  Their fundamental result is
given in their equation (A13) for the second moment of the
redistribution function including angular dependence.  Averaging that
result over angle depends on whether an isotropic or dipole phase
function is assumed.  The two results are, in terms of physical frequency,
\be
   D(\nu) = \varphi(\nu)+{1\over 3}\DeltaD^2\,\varphi''(\nu), \qquad 
     \hbox{(RD isotropic)},    \eqlab{18.3.3}
\ee
\be
 D(\nu) = \varphi(\nu)+{7 \over 20}\DeltaD^2\,\varphi''(\nu), \qquad 
     \hbox{(RD dipole)}.       \eqlab{18.3.4}
\ee
In fact, RD adopted a version without the second derivative,
\be
     D(\nu) = \varphi(\nu), \qquad \hbox{(RD adopted)}.
            \eqlab{18.3.5}
\ee
Their main rationale for this choice was that either the first or
second moment needed to be altered in order to preserve particle
conservation, and (\eqref{18.3.5}) was the simplest way of doing this.
Others \citep{MMR97,Chen04,Hirata05} have used the RD approximation
with this form for $D(\nu)$.

For much of formal developments in this paper, $D(\nu)$ will be left
general, and only later will particular choices be made.  One should
point out that all of the expressions
(\eqref{18.3.1})--(\eqref{18.3.5}) for $D(\nu)$ are asymptotically
equivalent in the line wings, so it only near the line core that
differences occur.  A desire for a non-singular result argues against
(\eqref{18.3.1}).  Although RD is somewhat ambiguous as to the correct
form, it seems clear that the correct cutoff for the asymptotic wing
region should occur on the scale of the Doppler width; this argues
against Basko's choice (\eqref{18.3.2}), which implies this scale is
the natural width.  In \S\ref{Kompaneets}, the choice for $D(\nu)$
will be discussed further.

The second category of F-P equations take recoil into
account.  This was done first by \citet{Basko81}, who gave the
following generalization of the F-P equation (\eqref{18}),
\be
  {1 \over c\chi}\pder{J}{t} =
   {\DeltaD^2\over 2} \pder{}{\nu} \left[ \Diff(\nu)
 \left(\pder{J(\nu)}{\nu} + {hJ \over kT} \right)\right], \eqlab{18.5}
\ee 
in which the new term $hJ/kT$ accounts for the recoil.  Basko used the
value (\eqref{18.3.2}) for $D(\nu)$.  RD introduced a slight variation
of Basko's generalized equation by using (\eqref{18.3.5}) for
$D(\nu)$, which was subsequently employed by CM.  This equation with
recoil will not be used directly in our derivation of a corrected F-P
equation, although we shall compare our results with it later.  Unless
indicated to the contrary, the term ``uncorrected F-P equation''
will refer strictly to equation (\eqref{18}).

\section{Detailed Balance and ``Corrected'' Redistribution}\label{detailed}

In order to describe correctly the approach to thermal equilibrium, the
redistribution function must satisfy a certain symmetry condition
known as detailed balance, which is a result of the microscopic reversibility
of the scattering process, along with the inclusion of appropriate phase space
factors and the fact that the scattering particles obey a Maxwellian
velocity distribution at temperature $T$.

\citet[][hereafter DW]{DW85} considered the detailed balance relations
satisfied by redistribution functions.  Their result was stated for
the redistribution function using energy based intensities, and they
also used opposite ordering of arguments $\nu$ and $\nu'$.  Accounting
for these differences, the DW result translates 
in our notation to
\be   
    R(\nu',\nu) \nu^2 e^{-h\nu / kT}
    =  R(\nu,\nu') {\nu'}^2 e^{-h\nu' / kT}. \eqlab{10}
\ee

In most cases of line transfer, one usually deals with frequencies
$\nu$ and $\nu'$ very close to the line center frequency and thus very
close to each other. Then equation (\eqref{10}) can be approximated by
the symmetry relation,
\be
   \Rstar(\nu',\nu) =  \Rstar(\nu,\nu').  \eqlab{13}
\ee
We have added the star to denote redistribution functions satisfying
the special symmetry (\eqref{13}) rather than (\eqref{10}).  In fact,
the usual redistribution functions used in line transfer calculations
are clearly approximate in that they satisfy (\eqref{13}) rather than
(\eqref{10}) \citep[e.g.,][]{Hummer62}.

Approximations such as equation (\eqref{13}) are satisfactory for many
applications.  However, as described in the Introduction,
there are cases where one needs a formulation that 
includes enough physics to ensure the detail balance relation (\eqref{10}).

     Let us consider how one may modify a simple redistribution
function obeying (\eqref{13}) so that it satisfies (\eqref{10}) instead.  
\citet{DW85} did this by choosing the
usual expression for $\nu > \nu'$, and obtaining the result for $\nu <
\nu'$ by the detailed balance relation [see discussion in RW after
equation (RW22)].  In our notation, this may be written,
\be
      R_{\rm DW}(\nu, \nu') 
           = \cases{ \Rstar(\nu, \nu'), &$\nu >\nu',$\cr
                     \Rstar(\nu', \nu)(\nu / \nu')^2
                     \exp[h(\nu'-\nu)/kT], &$\nu < \nu'$.\cr} \eqlab{16}
\ee
While this choice clearly will lead to thermal equilibrium in the
appropriate limit, it does have the undesirable property of creating a
discontinuity in slope in the redistribution function at $\nu=\nu'$,
even when one does not exist in $\Rstar$.

In this paper we use a different method of correction.
Again starting with an $\Rstar$ that satisfies the approximate detailed
balance relation (\eqref{13}), we define the ``corrected'' redistribution
function $\Rcorr$, which is related to the original $\Rstar$ through
a  ``diagonal similarity transformation,''
\be
     \Rcorr(\nu,\nu') = U(\nu)\Rstar(\nu,\nu') U^{-1}(\nu'), 
          \qquad {\rm or} \qquad
     \Rstar(\nu,\nu') = U^{-1}(\nu)\Rcorr(\nu,\nu') U(\nu'), 
\eqlab{17}
\ee
where,
\be
           U(\nu) = \nu e^{-{1 \over 2}h\nu/kT}. \eqlab{17.1}
\ee
It is easily verified that this corrected redistribution function
satisfies the detailed balance relation (\eqref{10}).   We prefer this
method to that of DW, because it introduces no new discontinuities of
slope at $\nu =\nu'$ and is more amenable to the kind of analytic
manipulations employed in this paper.

\section{Corrected Fokker-Planck Equation}

The correction proceedure given in equation (\eqref{17}) can be used
in any formulation of the redistribution problem.  For this paper,
however, we shall apply it to the Fokker-Planck equation (\eqref{18}).  In 
Appendix A, it is shown that the $R(\nu,\nu')$ implied by equation
(\eqref{18}) satisfies the symmetry condition (\eqref{13}), so that
equation (\eqref{17}) may be used to define a corrected redistribution
function. Using this, we straightforwardly derive our new corrected
Fokker-Planck equation that obeys detailed balance, namely,
\be
      {1 \over c\chi}\pder{J}{t} =  {\DeltaD^2 \over 2}
      \pder{}{\nu} \left\lbrace \Diff \left[ \pder{J}{\nu}
      + \left({h \over kT} -{2 \over \nu}\right) J + {c^2J^2 \over 2kT\nu^2}
     \right] \right\rbrace.  \eqlab{40.1}
\ee
(see Appendix B for the mathematical details of this derivation.)

We now identify the physical significance of the terms in the
inner set of brackets of this corrected F-P equation.  The first term,
$\partial J/\partial\nu$, is easily understood, since it is the same
diffusion term as in the uncorrected equation (\eqref{18}).  The last
term, involving the square of $J$, clearly describes the stimulated
scattering; let us ignore this term for the moment, since we want to
make comparisons with other F-P equations that have also ignored this
process; we shall return to it in \S\ref{Kompaneets}.

There are two remaining terms, $hJ/kT$ and $-J/2\nu$.  The first of
these is precisely the Basko recoil term, as in equation
(\eqref{18.5}).  This may seem remarkable, since the F-P
equation in its uncorrected form (\eqref{18}) does not include recoil at
all.  However, there is a simple physical explanation for this seeming
coincidence.  In the theory of Brownian motion, which is also
described by a F-P equation, Einstein found that thermal
equilibrium required a certain relations between the coefficients of
``diffusion'' and ``drift'' (other terms for drift are ``mobility''
or ``dynamical friction'').  In the present case, the uncorrected
equation (\eqref{18}) describes only the ``diffusion'' part of the process.
Imposing detailed balance through our correction procedure has
automatically generated the proper ``drift'' term, here given by the
recoil term.

The remaining term $-J/2\nu$ can be explained as follows.  With just
the diffusion term and the Basko recoil term, as in equation (\eqref{18.5}),
the equilibrium solution (setting the bracket
to zero) would give $J \propto \exp(-h\nu/kT)$.
In fact, \citet{Basko81}, \citep[following][]{Field59}, claimed that
his treatment of recoil led to thermal equilibrium because the
equilibrium solution of his equation was proportional to
$\exp(-h\nu/kT)$.  However, to be rigorously thermal (without
stimulated processes), $J$ should be proportional to the Wien law,
$\nu^2 \exp(-h\nu/kT)$, not just $\exp(-h\nu/kT)$.  The extra term
$-2J/\nu$ in the F-P equation (\eqref{40.1}) is just what is
needed to account for the missing factor of $\nu^2$.  In many
applications this additional slowly varying factor may be unimportant.
However, since the goal here is to account for thermal equilibrium as
precisely as possible, this term should be included as a matter of
principle.

\section{Relation to the Kompaneets Equation}\label{Kompaneets}

For further investigation of the corrected equation (\eqref{40.1}),
especially the stimulated scattering process, it is useful to write it
in terms of the photon occupation number $n$,
\be
      {1 \over c\chi}\pder{n}{t} = {\DeltaD^2 \over 2}{1 \over \nu^2}
      \pder{}{\nu} \left\lbrace \nu^2 \Diff \left[ \pder{n}{\nu}
      + {h \over kT}n(1+n) \right] \right\rbrace.  \eqlab{40}
\ee
This is reminiscent of the Kompaneets
equation \citep{Kompaneets57,RL79}, which can be written,
\be
      {1 \over N_{\rm e}\sigma_{\rm T}c}\pder{n}{t} 
    = {kT \over m_{\rm e}c^2}{1 \over \nu^2}
      \pder{}{\nu} \left\lbrace \nu^4 \left[ \pder{n}{\nu}
      + {h \over kT}n(1+n) \right] \right\rbrace,  \eqlab{40.5}
\ee
where $N_{\rm e}$ is the electron density
and $\sigma_{\rm T}$ is the Thomson cross section.

Some of the desirable general properties of the new equation follow
exactly as in the Kompaneets case.  It satifies photon conservation
exactly owing to the form of the right hand side as $1/\nu^2$ times a
frequency derivative.  It satisfies detailed balance by construction;
a result of this is that it has the correct thermal equilibrium
solution.  We can see this by setting the right hand side to zero,
which leads to the condition,
\be
 \pder{n}{\nu} + {h \over kT}n(1+n)=0.  \eqlab{44.1}
\ee
The general solution to this differential equation can be written
\be
     n = {1 \over e^{(h\nu-\mu)/kT} -1},  \eqlab{44.2}
\ee
where $\mu$ is a constant.  This is a thermal Bose-Einstein
distribution with chemical potential $\mu$, which is the correct
thermal distribution for photons obeying a conservation law.  We note
that the general properties of photon consevation and detailed balance
are independent of the particular form of $D(\nu)$.

The similarity between the corrected RD equation (\eqref{40}) and the
Kompaneets equation (\eqref{40.5}) can be emphasized further by using
definition (\eqref{18.2}) and choosing the RD value (\eqref{18.3.5})
for the diffusion coefficient $D(\nu)$.  Then equation (\eqref{40}) can be
written,
\be
      {1 \over c}\pder{n}{t} = {kT \over m_{\rm a}c^2} {1 \over \nu^2}
      \pder{}{\nu} \left\lbrace \nu^2\nu_0^2 \alpha \left[ \pder{n}{\nu}
      + {h \over kT}n(1+n) \right] \right\rbrace,  \eqlab{43}
\ee
where $m=m_{\rm a}$ is the atomic mass and $\alpha$ is
the line absorption coefficient,  
\be
        \alpha_{\rm L}(\nu) = \chi \varphi(\nu).  \eqlab{45}
\ee
We now make a minor modification of equation (\eqref{43}) by replacing
$\nu_0^2$ by $\nu^2$.  This is a very slight change, since
$\alpha(\nu)$ is significant only in a narrow region around line
center, so the distinction between $\nu$ and $\nu_0$ is slight.  Such
a replacement is not unusual for line problems; it is quite typical
when deriving line redistribution functions, and is even found in the
definition of the Doppler width, equation (\eqref{18.2}).  

One may also regard the above modification as equivalent to a
slightly different diffusion coefficient $D(\nu)$, and, as noted
above, the important properties of photon conservation and detailed
balance are independent of $D(\nu)$.  With this modification, equation
(\eqref{43}) becomes
\be
      {1 \over c}\pder{n}{t} = {kT \over m_{\rm a}c^2} {1 \over \nu^2}
      \pder{}{\nu} \left\lbrace \nu^4 \alpha \left[ \pder{n}{\nu}
      + {h \over kT}n(1+n) \right] \right\rbrace.  \eqlab{44}
\ee

Remarkably, equation (\eqref{44}) is precisely the same as the
Kompaneets equation if we take $m$ to be the electron mass $m_{\rm e}$
and $\alpha$ to be the Thomson scattering coefficient, $\alpha_{\rm
T}=N_{\rm e}\sigma_{\rm T}$.  Thus, equation (\eqref{44}) provides a
{\em unification} of the Fokker-Planck equation for resonance line scattering
and the Kompaneets equation.  One only needs to substitute the
appropriate scattering coefficient and mass of the scatterer.

It should be noted that this unification is even valid if one chooses
to use one of the forms (\eqref{18.3.3}) or (\eqref{18.3.4}) for
$D(\nu)$ instead of (\eqref{18.3.5}).  In either case, this simply
involves replacing $\alpha$ by $\alpha + C\alpha''$ in equation
(\eqref{44}), where $C$ is a constant.  This will still be completely
consistent with the Kompaneets equation, because $\alpha$ for Thomson
scattering is independent of frequency.  However, for the remainder of
this paper we shall continue to use the corrected Fokker-Planck
equation in the simple form (\eqref{44}).

\section{Energy Exchange Formula}\label{exchange}

Equation (\eqref{44}) can be used to derive an energy exchange formula
between the radiation and kinetic energy of the atoms.  The energy density
of the radiation field is
\be
     U= {8\pi h \over c^3} \int \nu^3 n\, d\nu.  \eqlab{47}
\ee
Differentiating with respect to time and using equation (\eqref{44}),
followed by an integration by parts, yields the rate of change of radiation
energy density,
\be
     \pder{U}{t} = -{8\pi h k T \over m_{\rm a} c^4} 
                 \int \nu^4 \alpha_{\rm L}(\nu)\left[ \pder{n}{\nu}
      + {h \over kT}n(1+n) \right]d\nu,  \eqlab{48}
\ee
which is the negative of the heating rate for the gas.

Equation (\eqref{48}) is the general formula for energy exchange
within our Fokker-Planck formulation.  The presence of the frequency
derivative in the integral makes the formula difficult to interpret,
but it may be put into a more revealing form by defining at each
frequency an appropriate temperature of the radiation field by
comparison to a thermal Bose-Einstein distribution with temperature
$T$ and chemical potential $\mu$,
\be
        n_{\rm th}= {1 \over e^{(h\nu-\mu)/kT} -1} 
   \qquad\hbox{for which}\qquad  n_{\rm th}'=-{h \over kT}n(1+n).
\eqlab{50}
\ee
Given the values at frequency $\nu$ of an arbitrary radiation
field $n(\nu)$ and its slope $n'(\nu)$, equation (\eqref{50}) one may
define the local ``radiation temperature'' $\TR(\nu)$ for an arbitrary
radiation field by
\be
       \TR(\nu) = -{h \over k} {n(1+n) \over n'}. \eqlab{51}
\ee 
With this definition, equation (\eqref{48}) becomes
\be
     \pder{U}{t} = -{8\pi h^2 \over m_{\rm a} c^4} 
                 \int \nu^4 \alpha_{\rm L}(\nu) n(1+n)
              \left( 1-{T \over \TR(\nu)} \right) d\nu.  \eqlab{52}
\ee

In order to interpret this equation, let us consider the special case
when the radiation field is smooth and slowly varying in the
neighborhood of the line center at $\nu_0$, so it and its derivative
can be considered constant.  Then the delta-function character of
$\alpha_{\rm L}(\nu)$ implies,
\be
    \pder{U}{t} = -{8\pi h k T \chi \nu_0^4 \over m_{\rm a} c^4}
              n_0(1+n_0) \left( 1-{T \over T_{\rm R0}} \right),  \eqlab{49}
\ee
where the subscript ``0'' means evaluation at $\nu_0$.

This may be put into another form by defining the rate of
scattering per unit volume by
\be
      P= 4\pi \int    \alpha_{\rm L} (1+n)J\, d\nu =
   {8\pi \chi \over c^2} \nu_0^2 n_0(1+n_0),
    \eqlab{53}
\ee
where we have again used the delta-function character of $\alpha_{\rm
L}(\nu)$.  Note that stimulated scatterings are included in this
rate. Then our energy exchange formula takes the simpler form,
\be
   \pder{U}{t}=-P \left({h\nu_0 \over m_{\rm a} c^2}\right) h\nu_0 
        \left( 1- {T \over T_{\rm R0}} \right).    \eqlab{54}
\ee
(We remark that the general energy exchange formula (\eqref{52}) could be
written in this same simplified form, provided one uses for the radiation
temperature a suitable mean value, frequency-averaged over the line.)

The factor $(1-T/T_{\rm R0})$ in the formula (\eqref{54}) shows how
the energy exchange between the radiation and the matter depends on
their relative temperatures.  In particular, energy always flows from
the hotter to the colder system; no energy exchange occurs when the
radiation and matter temperatures are the same.  These fundamental
principles of thermodynamics are guaranteed by the detailed balance
built into our equations.

The factor $(1-T/T_{\rm R0})$ is particularly important for situations
with large optical depth in the resonance line. In these cases, the
Wouthuysen-Field effect \citep{Wouthuysen52,Field59} becomes relevant.
According to this effect, the scattering within the line changes the
local radiation field in such a way that the radiation temperature
within the line is brought close to the gas temperature.  Therefore the factor
$(1-T/T_{\rm R0})$ becomes very small and so greatly reduces the
flow of energy between the radiation and the gas, to a value much less
than one what might estimate from simple arguments based on the
radiation temperature of any incident radiation.

We are now in a position to compare our energy exchange formulas to
those of \citet[][hereafter MMR]{MMR97} and \citet[][hereafter
CM]{Chen04}.

\subsection{Comparison with MMR}

In deriving their energy exchange formula, MMR assumed (implicitly)
the near constancy of the radiation field near line center, so it
should be compared with our equation (\eqref{54}).  Some translation
is necessary, since their rates are defined per atom instead of per
unit volume. Making that adjustment, the MMR formula (their
Eq.\ [30]) is, in our notation,
\be
   \pder{U}{t}=-{\tilde P} \left({h\nu_0 \over m_{\rm a} c^2}\right) h\nu_0.
          \eqlab{54.1}
\ee
The rate ${\tilde P}$ is the same as our rate $P$, defined in equation
(\eqref{53}), except that it omits the stimulated scattering factor
$(1+n)$.

The omission of stimulated scattering is minor, but more importantly
the MMR formula does not contain the factor $(1-T/T_{\rm R0})$, which,
as we have seen, describes how the energy exchange depends on the
relative temperatures of the matter and radiation.  The MMR result
will be valid only when the radiation temperature is much greater than
the matter temperature, so that the recoil effect dominates the
counterbalancing diffusive effect of the thermal electrons.

In Appendix B of their paper, MMR outlined a derivation of their
energy exchange formula from a Fokker-Planck equation that contained
the Basko recoil term.  In that derivation they eventually dropped the
diffusion term and kept only the recoil terms.  They clearly recognized 
that because of this their formula would only apply for photons of
much higher energies than the thermal energies of the atoms.  

MMR were reluctant to include the diffusion term, because they argued
that only terms of order $(v/c)$ had been included and that one needed
to include terms up to $(v/c)^2$ in order to get the reverse flow of
energy from the atoms to the radiation.  However, the present approach
using detailed balance automatically provides the correct description of the
reverse flow, as evidenced by the correct thermal equilibrium.

\subsection{Comparison with CM}

Another approach to energy exchange was given by \citet[][hereafter
CM]{Chen04}, who used the F-P equation with Basko recoil term
(\eqref{18.5}) to find the energy exchange rate in a particular
cosmological context.  By using this equation, CM assumed (as MMR did
not) that this equation could be relied upon to describe the reverse
flow of energy.  

CM used numerical solutions to equation (\eqref{18.5}) to evaluate the
energy transfer and did not derive a general formula such as our
equation (\eqref{54}), but we may do so now.  In a sense the present
derivation is merely extending the derivation outlined by MMR in their
Appendix B, except that here the diffusion term is included.

Multiplication of equation (\eqref{18.5}) by $h\nu$ and integrating
over $\nu$ gives, after integration by parts and other straightforward
manipulations,
\be
      \pder{U}{t} = -{4\pi h k T \over m_{\rm a} c^2} 
                 \int \nu_0^2 \alpha_{\rm L}(\nu)\left[ \pder{J}{\nu}
      + {h J\over kT} \right]d\nu.   \eqlab{55}
\ee
In order to compare this with our result (\eqref{52}), we define a
new ``radiation temperature'' $T_{\rm CM}$ by the relation
\be
      T_{\rm CM} = -{h \over k} {J \over J'}. \eqlab{56}
\ee 
This differs from the definition (\eqref{50}) by ignoring stimulated
emission and also the factor $\nu^2$ in the relation between $n$ and
$J$.  It is equivalent to assuming that a thermal radiation field is
proportional simply to $\exp(-h\nu/kT)$.

Under the same assumptions as made for equation (\eqref{54}), we obtain
the following energy exchange formula,
\be
   \pder{U}{t}=-{\tilde P} \left({h\nu_0 \over m_{\rm a} c^2}\right) h\nu_0 
 \left( 1- {T \over T_{\rm CM0}} \right),  
    \qquad \hbox{(CM).}  \eqlab{57}
\ee
We shall call this the ``CM formula,'' because it is a natural
consequence of their approach, even though they did not derive it
explicitly.  The CM formula differs from for our equation (\eqref{54})
by the replacement of $T_{\rm R0}$ by $T_{\rm CM0}$ (and the fact that
${\tilde P}$ does not include stimulated scatterings).

The most noteworthy feature of the CM formula is that it shares with
(\eqref{54}) the desirable property of taking into account the
relationship between matter and radiation temperature (although with
an slightly approximate radiation temperature).  In particular, the
approach of CM does take into account of the suppression of energy
transfer for lines of high optical depth, in accordance with the
Wouthuysen-Field effect.  This clearly explains why CM found much
lower energy exchange rates than MMR.

Let us make some estimate of the difference between the radiation
temperatures $\TR$ and $T_{\rm CM}$.  Stimulated emission is usually
fairly negligible for resonance line transfer, so let us ignore it for
the moment.  Then one can easily show that,
\be
       T_{\rm CM} = {\TR \over 1-2k\TR /h\nu}.  \eqlab{56.1}
\ee

As an practical example, in problems involving hydrogen Lyman alpha,
gas temperatures are typically below $\sim 10^4$ K, since anything
much higher leads to rapid ionization of hydrogen.  The equivalent
temperature for the Lyman alpha transition is $\sim 10^5$ K, so by
equation (\eqref{56.1}) the two radiation temperatures $\TR$ and
$T_{\rm CM}$ may differ by up to $\sim 20$\%.  However, for lower
radiation temperatures, they will differ proportionally less, and the
CM radiation temperature $T_{\rm CM}$ may be a reasonable
approximation to the more accurate $\TR$.  For very sensitive
applications, however, it would be prudent to use the accurate form.

\section{Discussion}

A new form for the Fokker-Planck equation desribing resonant line
scattering has been derived that incorporates detailed balance and
stimulated scattering.  When put into the form (\eqref{44}) the new
equation shows a surprising unification with the Kompaneets equation.

The new Fokker-Planck equation has been used to derive expressions for
the rate of energy transfer between the radiation and the thermal gas,
incorporating both recoil and the proper reverse flow of energy
from the gas to the radiation.  We have compared our energy exchange
result to those used by \citet{MMR97} and \citet{Chen04}. 

In their paper MMR noted that their energy exchange formula would be
limited to cases where the radiation temperature greatly exceeds the
gas temperature.  We have verified this explicly by deriving the
general dependence on the radiation and gas temperatures, as seen,
e.g., in the factor $(1-T/T_{\rm R0}$ in our equation (\eqref{54}).
This factor is thermodynamically required, and also will be absolutely
crucial in cases of resonance line scattering of high optical depth,
where the Wouthuysen-Field effect can play a pivotal role.

We have also derived an approximate explicit energy
transfer formula closely following the work of CM, which
is based on the Fokker-Planck equation (\eqref{18.5}).  The resulting
energy exchange formula (\eqref{57}) (which we call the CM formula) does
approximately incorporate detailed balance and does approximately
account for the relative radiation and gas temperatures.  The CM
formula should give acceptable results in most cases of interest.
This demonstrates that the CM approach to energy transfer is clearly
superior to the more limit formula of MMR.

However, although the Fokker-Planck equation (\eqref{18.5}) and the
associated CM energy exchange formula would seem to provide adequate
results for most situations, there may be other reasons for preferring
our new formulation of the Fokker-Planck equation (\eqref{44}).  From
a purely theoretical point of view it treats the thermodynamics in a
more precise manner, and has the esthetically pleasing property of
being unified in form with the Kompaneets equation.  There may also be
practical reasons for the using our more accurate formulation.  For
example, in determining the spin temperature of the 21cm line in
cosmological contexts, one needs to consider the delicate balance of
different processes, and here a thermodynamically correct description
of Lyman alpha scattering could prove to be a crucial element.  We
plan to investigate this in the near future.

The results of the present paper suggest some avenues for possible
future work.  The unification of the resonance line and electron
scattering Fokker-Planck equations discussed in \S\ref{Kompaneets}
strongly suggests that one might develop a single derivation of both
equations, based on expansions of the transfer equations with
redistribution functions that take all physical effects explicitly
into account.  Done carefully, this could resolve the somewhat
unfortunate ambiguity of what the proper value of $D(\nu)$ should be.

One limitation of the present paper is its assumption of a homogeneous
and isotropic radiation field $J(\nu)$, a limitation shared by the
Kompaneets equation.  It would be desirable to derive a Fokker-Planck
equation for resonance scattering with a spatially dependent and
anisotropic radiation field, $I(\nu,\bfr,\bfn)$, dependent on position
$\bfr$ and direction $\bfn$.  This would be useful for cases where the
incident irradiation is anisotropic, or in cases where the medium
itself is inhomogeneous.

There is one way to obtain an approximate equation for anisotropic
radiation fields, one that is commonly used with the Kompaneets
equation. Neglecting stimulated emission, we write,
\be
    {1 \over c}\pder{I}{t} +\bfn\cdot\nabla I = -\chi \phi I +\chi\phi  J
 +\chi {\DeltaD^2 \over 2} \pder{}{\nu} \left\lbrace \Diff \left[ \pder{J}{\nu}
 + \left({h \over kT} -{2 \over \nu}\right) J \right] \right\rbrace.  \eqlab{58}
\ee       
This is equivalent to assuming that the extinction part of the
scattering is treated with the directional intensity, but the emission
part is treated as if the radiation field were isotropic, as in
equation (\eqref{40.1}).  Since scattered radiation is not too far
from being isotropic, and becomes more so after many scatterings, this
is often an acceptable approximation.  We note that when the radiation
field is homogeneous and isotropic, (\eqref{58}) reduces to equation
(\eqref{40.1}) when integrated over all solid angles.

However, equation (\eqref{58}) does have some definite defects.  If
one wanted to include stimuated scattering, the extinction term would
have to depend on angle in some as-yet unknown way; this defect is not too
serious, since stimulated scattering is not likely to be important for
any resonance lines.  A more serious defect is that the equation does
not include the correlation of the frequency diffusion and shift on the
scattering angle, an effect we know is very strong from the form of
the angle-dependent redistribution function $R_{II}(\nu,\bfn;\nu',\bfn')$
\citep{Hummer62}.

Therefore, for theoretical reasons, as well as practical, it would be
desirable to derive a Fokker-Planck equation for resonance scattering
with a fully anisotropic radiation field.  This could be done using
the same correction procedure of the present paper, but now applied to
the angle-dependent results of RD, their equation (A16).  Such a
derivation is currently under investigation.

\acknowledgements
The author would like to thank Avi Loeb for helpful discussions and comments.

\appendix

\section{A. Symmetry of the Fokker-Planck Redistribution Function}

The redistribution function from which the RD Fokker-Planck equation
(\eqref{18}) was derived satisfies the symmetry
$R(\nu,\nu')=R(\nu',\nu)$.  However, because approximations have been
made, it is not obvious that the Fokker-Planck redistribution function
$R$ defined implicitly through equation (\eqref{18}) must necessarily
satisfy this same symmetry.  To investigate this it is useful to
consider the following double integral,
\be
     \int\int G(\nu)R(\nu,\nu') F(\nu') \,d\nu\,d\nu'. \eqlab{A1}
\ee
This quantity is bilinear in the two functions $F(\nu)$ and $G(\nu)$,
which are completely arbitrary, except that they and their derivatives
are assumed to vanish sufficiently rapidly at the limits of
integration $(0,\infty)$.

We write the basic RD F-P result (\eqref{5}) for an
 arbitrary function $F(\nu)$,
\be
 \int R(\nu,\nu') F(\nu')\,d\nu'  = \varphi(\nu) F(\nu) 
 +{\DeltaD^2\over 2} \pder{}{\nu}
    \left[ \Diff(\nu)  \pder{F(\nu) }{\nu} \right]. \eqlab{A2}
\ee 
Using this for the integral over $\nu'$ in (\eqref{A1}), we have
\be
   \int\int G(\nu)R(\nu,\nu') F(\nu') \,d\nu\,d\nu' 
    = \int G(\nu)\left\lbrace \varphi(\nu)F(\nu) 
  +{\DeltaD^2\over 2}\pder{}{\nu} \left[ \Diff(\nu)\pder{F(\nu)}{\nu} 
      \right] \right\rbrace\,d\nu.  \eqlab{A3}
\ee
Integrating by parts twice, assuming zero 
contributions at the limits, gives
\be
    \int\int G(\nu')R(\nu',\nu) F(\nu) \,d\nu\,d\nu'
       =\int \left\lbrace \varphi(\nu)G(\nu)
 +{\DeltaD^2\over 2}\pder{}{\nu} \left[ \Diff(\nu)\pder{G(\nu)}{\nu} 
     \right] \right\rbrace F(\nu)\,d\nu,  \eqlab{A4}
\ee
where we have interchanged $\nu \leftrightarrow \nu'$ in the double 
integral on the left hand side. 
Interchanging $F \leftrightarrow G$ yields
\be
   \int\int G(\nu)R(\nu',\nu) F(\nu') \,d\nu\,d\nu' 
    = \int G(\nu)\left\lbrace \varphi(\nu)F(\nu) 
  +{\DeltaD^2\over 2}\pder{}{\nu} \left[ \Diff(\nu)\pder{F(\nu)}{\nu} 
      \right] \right\rbrace\,d\nu.  \eqlab{A5}
\ee
Subtracting this equation from (\eqref{A3}) gives
\be
 \int\int G(\nu)\left[R(\nu,\nu') -R(\nu',\nu) \right]
          F(\nu') \,d\nu\,d\nu' = 0.  \eqlab{A6}
\ee
This relation is true for arbitrary $G(\nu)$ and $F(\nu)$, which implies,
\be
            R(\nu,\nu') =R(\nu',\nu), \eqlab{A7}
\ee
that is, the RD F-P approximation maintains the exact symmetry of the
redistribution function on which it was based.
	
\section{B. Derivation of the ``Corrected'' Fokker-Planck Equation}

Since $R(\nu,\nu')$ is symmetric, we may write for for an arbitrary
function $G(\nu)$,
\be
       \int G(\nu')R(\nu',\nu) \,d\nu' = \varphi(\nu)G(\nu)
     +{\DeltaD^2\over 2}\pder{}{\nu} 
    \left[ \Diff(\nu)\pder{G(\nu)}{\nu} \right].\eqlab{B1}
\ee
The symmetry also allows us to use equation (\eqref{17}) to give
``corrected'' versions of equations (\eqref{A2}) and (\eqref{B1}),
namely,
\bea
     \int \Rcorr(\nu,\nu') F(\nu')\,d\nu' &=& \varphi F 
+{\DeltaD^2\over 2} U\pder{}{\nu}\left[ \Diff \pder{}{\nu}
         \left( U^{-1}F \right) \right],
       \eqlab{B2}\\
     \int G(\nu')\Rcorr(\nu',\nu) \,d\nu' &=& \varphi G 
+{\DeltaD^2\over 2} U^{-1} \pder{}{\nu}\left[ \Diff \pder{}{\nu}
         \left( UG \right) \right],
       \eqlab{B3}
\eea
for arbitrary functions $F(\nu)$ and $G(\nu)$. 

Based on the preceding, we now derive a corrected F-P
transfer equation.  Stimulated emission can be easily incorporated
into this derivation by starting with the full equation,
\be
   {1 \over c\chi}\pder{J}{t} = \int \left\lbrace 
  \left[ 1+ n(\nu) \right]\Rcorr(\nu,\nu')J(\nu')
 -\left[ 1+ n(\nu') \right]\Rcorr(\nu',\nu)J(\nu)
     \right\rbrace d\nu'.    \eqlab{B4}
\ee
The integrals over $\nu'$ can be performed using 
(\eqref{B2}) with $F=J$ and (\eqref{B3}) with $G=1+n$, which gives,
\be
      {1 \over c\chi}\pder{J}{t} = {\DeltaD^2 \over 2}
    \left\lbrace
      ( 1+ n)    U\pder{}{\nu}
      \left[ \Diff \pder{}{\nu}
         \left( U^{-1} J \right) \right]  
   -  U^{-1} J \pder{}{\nu}
     \left[ \Diff \pder{}{\nu}\left( ( 1+ n)U \right) \right] 
     \right\rbrace. \eqlab{B5}
\ee
The right hand side can be written as a frequency derivative,
\be
      {1 \over c\chi}\pder{J}{t} = {\DeltaD^2 \over 2}
    \pder{}{\nu} \left\lbrace (1+n)U\Diff\pder{}{\nu} \left( U^{-1} J \right)
   -\left( U^{-1} J \right)
    \Diff\pder{}{\nu}\left[(1+n)U\right] \right\rbrace.  \eqlab{B6}
\ee
Substituting $n=c^2J/2\nu^2$ and using equation (\eqref{17.1}), we
find, after some straightforward manipulations, the ``corrected''
equation (\eqref{40.1}) in terms of the mean intensity.

\end{document}